\documentclass[%
reprint,
 amsmath,amssymb,
 aps,
]{revtex4-2}

\usepackage{graphicx}
\usepackage{dcolumn}
\usepackage{bm}

\begin{document}

\title{Kilo-Tesla axial magnetic field generation with high intensity spin and orbital angular momentum beams}

\author{Andrew Longman}
 \affiliation{Lawrence Livermore National Laboratory, Livermore, California 94551}
\email{longman1@llnl.gov}
\author{Robert Fedosejevs}%
 \affiliation{Department of Electrical and Computer Engineering, University of Alberta, Edmonton T6G1R1, Canada}%

\date{\today}

\begin{abstract}
Absorption of angular momentum from a high intensity laser pulse can lead to the generation of strong axial magnetic fields in plasma. The effect, known as the inverse Faraday effect can generate kilo-Tesla strength, multi-picosecond, axial magnetic fields extending over hundreds of microns in underdense plasma. In this paper we explore the effect with ultra-high intensity circularly polarized Gaussian beams and linearly polarized orbital angular momentum beams comparing analytic expressions with 3D particle-in-cell simulations. We develop a model for the transverse magnetic field profiles, introduce a new model for the temporal decay, and show that while the magnetic field strength is independent of plasma density, it has a strong dependence on the laser beam waist. 

\end{abstract}

\maketitle


\section{Introduction}



The inverse Faraday effect (IFE) describes the generation of axial magnetic fields when angular momentum is transferred from a laser pulse to plasma. If the pulse is circularly polarized (CP) and of relativistic intensity, the generated magnetic field strength can be on the order of 100’s of Tesla, last several picoseconds, and extend over millimeter scales \cite{najmudin01}. The IFE has been studied extensively with CP beams both theoretically \cite{pitaev61,haines01,naseri10,frolov04,bychenkov_tikhonchuk_1996,sheng96,Lecz16}, and experimentally \cite{najmudin01,deschamps70,Horovitz97}; yet the effect is not exclusive to CP beams but also applicable to lasers carrying orbital angular momentum (OAM) \cite{Ali10,Nuter18,Nuter20}. Currently, we only know of one analytic model describing OAM driven magnetic fields and its transverse profile \cite{Ali10}, but has not been verified numerically or experimentally. Furthermore, little is known about the axial extent of the magnetic fields or their lifetime, for both CP and OAM drivers.

There are several advantages of using OAM beams in high powered lasers as opposed to CP beams: unlike CP beams that carry a maximum spin angular momentum of $\sigma_z\hbar$ per photon where $\sigma_z=\{\pm 1\}$ is the spin number ($\sigma_z=0$ for linearly polarized (LP) beams), OAM beams carry an angular momentum of $(\sigma_z+\ell)\hbar$ per photon where $\ell$ is an azimuthal mode number that can take any integer value \cite{Allen92}. OAM has also been shown to couple more efficiently to free electrons due to an optimal overlap between the laser intensity profile and the angular momentum density yielding greater coupling at larger radial distances \cite{TIKHONCHUK2020}. Additionally, high intensity OAM beams have recently been produced in high power lasers using off-axis spiral phase mirrors mitigating nonlinear effects like temporal distortions and B-integral present with transmissive optics such as quarter waveplates \cite{Longman:20}.    

Recent works have looked at simulating IFE driven magnetic fields from OAM beams in various configurations: OAM beams with radial and azimuthal polarization \cite{Nuter18}, amplification of seeded magnetic fields \cite{Wu_2017}, spatiotemporal light springs \cite{Shi18}, and most recently linearly polarized OAM beams \cite{Nuter20}. In these studies the laser intensities were of moderate intensity $(I_0\approx 10^{18}$ Wcm$^{-2})$ verifying the existence of weaker magnetic fields $(\approx 10$ T$)$, with little modelling of the spatial or temporal properties of the magnetic fields.

In this work, we explore the spatial, and for the first time the temporal scaling of IFE magnetic fields driven by ultra-high intensity $(I_0\approx 10^{20}$ Wcm$^{-2})$ CP Gaussian and LP OAM beams. We develop new analytic models for both the spatial and temporal scales of the magnetic field and verify them with full 3D relativistic particle-in-cell (PIC) simulations. We demonstrate strong coupling of laser angular momentum to plasma through ponderomotive forces, resulting in axial magnetic fields in excess of 1 kT, more than 200 $\mu$m in length, and lasting several picoseconds in good agreement with the predictions of our analytic models, and previous experiments using CP Gaussian beams. 

\section{Coupling of high intensity angular momentum to plasma}

A circularly polarized beam is classically defined by its polarization vector rotating about the laser propagation axis $k_z$ in either a clockwise or counter-clockwise direction. Similarly, a beam carrying a well defined OAM is typically characterized by a helical wavefront also orientated about the $k_z$ vector with the helix rotating in either a clockwise or counter-clockwise direction \cite{Allen92}. Unlike a Gaussian beam, paraxial focusing of an OAM beam produces a complicated field structure described by modified-Bessel or hypergeometric functions but can be modelled using a suitably chosen basis set such as the Laguerre-Gaussian (LG) modes \cite{Longman:20b}. The electric field magnitude of a general OAM laser pulse can then be approximately given by $E=E_0 \psi_\ell g(t) \exp[i(\omega_0 t-k_z z)]$ where $E_0$ is the peak electric field strength of the fundamental Gaussian mode ($\ell=0$), $g(t)$ is the temporal envelope, and $\omega_0$ is the laser frequency. $\psi_\ell$ are the azimuthal LG modes (zero-radial mode) and can be given in the following compact form \cite{Allen92,BARNETT1994},
\begin{equation} \label{eq1}
    \psi_\ell=\frac{1}{\sqrt{|\ell|!}}\left(\frac{r\sqrt{2}}{w_0}\right)^{|\ell|}\left(\frac{z_0}{Z}\right)^{|\ell|+1}exp\left(-\frac{kr^2}{2Z}\right)e^{i\ell\phi},
\end{equation}
where $w_0$ is the beam waist in the focal plane, $z_0=kw_0^2/2$ is the Rayleigh range, and $Z=z_0+iz$ is the complex beam parameter. Taking the modulus squared of the field amplitude given in Eq. \ref{eq1} for $|\ell|\geq 1$, we find the so called “donut” mode intensity profiles that are symmetric about the $k_z$ axis; the helical phase is lost in the modulus resulting in no azimuthal structure in the intensity profile \cite{Allen92}. Increasing the azimuthal mode integer $|\ell|$ increases the OAM in the beam, increases the donut radius and by conservation of energy, the peak intensity around the donut decreases for a fixed laser energy and focusing geometry \cite{Longman:17}. As the peak intensity varies for different values of $|\ell|$ in a non-trivial way, we opt to use the beam power instead as it is constant for all $|\ell|$ modes.    

The ponderomotive force acting on an electron associated with a high intensity focal spot can be given by, $\bm F_p=-m_e c^2 \bm\nabla\sqrt{1+|\bm{a}|^2}$ where $\bm{a}=e\bm{E}/m_e c \omega_0$ is the normalized vector potential, $e$ and $m_e$ are the electron charge and mass respectively, and $c$ is the speed of light. In the standard case of a high-intensity Gaussian beam a finite energy maybe transferred from the laser to plasma electrons through ponderomotive scattering \cite{Mackenroth_2019}. The same is true for higher order LG modes \cite{Akou20}, with some possibility for electron trapping and additional energy transfer within the donut mode leading to a possible increased absorption rate \cite{Shi21,Miyazaki_2005}. However, ponderomotive scattering to first order is an intensity driven phenomenon, and no OAM is transferred to the plasma as the LG mode intensity has no azimuthal component. 

Recent work has analyzed the motion of free electrons in a LG beam in more detail, expanding to first and second order perturbations \cite{Nuter20,TIKHONCHUK2020}. It was shown that angular momentum can be transferred to free electrons in LP LG beams when considering second order terms in the equations of motion; the electron gains and loses angular momentum but averaging over one laser cycle yields zero net angular momentum transfer. An asymmetry in the laser field such as the ponderomotive force, or collisions are therefore needed to transfer a net OAM to the electrons. 

\section{The Inverse Faraday Effect}

To model the IFE, we start with the angular momentum conservation model used in Haines \cite{haines01}. In this model, coupling of the angular momentum density in the plasma with that of a driving laser is performed through an arbitrary absorption parameter yielding the conservation equation, 
\begin{eqnarray}\label{eq1.5}
&&n_e m_e r \frac{dv_{e\theta}}{dt}=-n_eerE_\theta \nonumber \\
&&-n_eer(v_{ez}B_r-v_{er}B_z)+\alpha_{abs}M_zc-n_em_e\nu_{ei}rv_{e\theta}.
\end{eqnarray}
Here, $n_e$ is the electron density, $\alpha_{abs}$ is the laser absorption fraction per unit length, $M_z$ is the laser angular momentum density, $v_{er}, v_{e\theta}$ are the radial and azimuthal electron velocities respectively, and $\nu_{ei}$ is the electron-ion collision frequency. $E_\theta, B_r,$ and $B_z$ are the plasma azimuthal electric, radial magnetic and axial magnetic fields respectively. Assuming the electrons are in steady state, ignoring the inertial and collisional terms, and using Faraday's law, we obtain the time rate of change of the axial magnetic field approximately as \cite{haines01}, 
\begin{equation} \label{eq2}
    \frac{\partial B_z}{\partial t} \approx\frac{c}{re}\frac{\partial}{\partial r}\frac{\alpha_{abs} M_z}{n_e} ,
\end{equation}
For an OAM mode that is either CP or LP, the angular momentum density can be given by \cite{Ali10,Allen92},
\begin{equation}\label{eq2b}
    M_z = \frac{I_0|\psi_\ell|^2g(t)^2}{\omega_0 c}\left[\ell+\sigma_z\left(|\ell|-\frac{2r^2}{w_0^2}\right)\right].
\end{equation}
Radial differentiation of Eq. \ref{eq2b} is straightforward, and if we assume a Gaussian temporal function, then the integration of Eq. \ref{eq2} is trivial given $\int g(t)^2dt\approx3\tau/4$ where $\tau$ is the temporal full-width half maximum (FWHM). By introducing the pulse temporal function $g(t)$, we are limiting the IFE to only be driven when the laser field is present. Though the magnetic field does not instantaneously disappear after the laser has passed, its evolution is no longer governed by the IFE, but rather the plasma dynamics which will be discussed later. 


Substituting Eq. \ref{eq1} into Eq's. \ref{eq2} and \ref{eq2b}, assuming the laser absorption rate and electron density is constant in space and time (no ponderomotive channelling), a Gaussian temporal pulse shape, and substituting the laser power $P=\pi I_0w_0^2/2$ we obtain the following expression for the axial magnetic field, 
\begin{eqnarray}\label{eq3}
&&B_z=\frac{3\alpha_{abs}P\tau}{\pi e n_e \omega_0 w_0^4}|\psi_\ell|^2 \times \nonumber \\
&&\left[\ell\left(\frac{|\ell|w_0^2}{r^2}-2\right)+\sigma_z\left(\frac{|\ell|^2w_0^2}{r^2}-4|\ell|+\frac{4r^2}{w_0^2}-2\right)\right].
\end{eqnarray} 

\begin{figure} 
\includegraphics{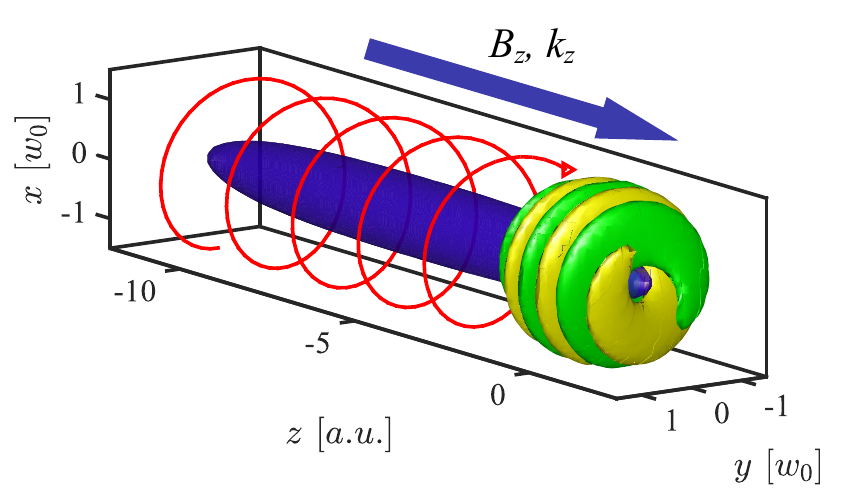}
\caption{\label{fig1} Illustration of an IFE driven B-field with a LP $\ell=1$ OAM mode and the corresponding electron helical trajectory in red with an arrow head showing the velocity direction. The green and yellow isosurfaces show the positive and negative electric fields of the laser respectively while the blue isosurface shows the positive magnetic field direction as indicated by the arrow.}
\end{figure}

The direct proportionality on $\ell$ and $\sigma_z$ in Eq. \ref{eq3} indicates that the axial magnetic field direction can be controlled by changing the sign of OAM helicity, or the CP handedness. The helicity of a LP OAM mode, the magnetic field direction and the corresponding electron motion is illustrated in Fig. \ref{fig1}. 

\begin{figure} 
\includegraphics{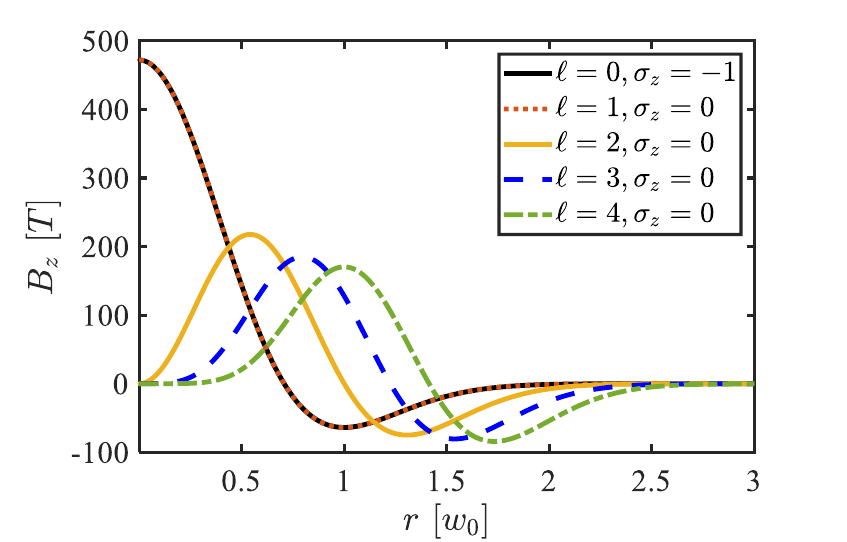}
\caption{\label{fig2} Transverse IFE driven magnetic field profiles driven from a CP Gaussian beam and various LP OAM beams given by Eq. \ref{eq3} and the legend. The following laser plasma parameters were used: $n_e$=0.03$n_c$, $\lambda=1$ $\mu$m, $P=65$ TW, $w_0=6$ $\mu$m, $\alpha_{abs}=0.435$ mm$^{-1}$, and $\tau=100$ fs.}
\end{figure}

Several transverse magnetic field profiles are plotted in Fig. \ref{fig2} with laser plasma parameters: $n_e=0.03n_c$, $\lambda=1$ $\mu$m, $w_0=6$ $\mu$m, $\alpha_{abs}=0.435$ mm$^{-1}$, $\tau$=100 fs, and $P=$ 65 TW, where $\lambda$ is the laser wavelength, and $n_c$ is the critical plasma density. We note that the $(|\sigma_z|=1,|\ell|=0)$ and $(\sigma_z=0, |\ell|=1)$ modes produce the same magnetic field profiles peaked on axis, while the higher order $\ell$ modes produce more complex coaxial structures that are zero on axis. Given the magnetic field peak on axis and the higher magnetic field strength, the $(|\sigma_z|=1,|\ell|=0)$ and $(\sigma_z=0, |\ell|=1)$ driven fields are probably more desirable for most applications. For the purpose of this work, we do not consider beams that are both circularly polarized and have OAM, but we remark that a beam with $(\sigma_z=\pm 1,\ell=\pm 1)$ produces the strongest on-axis magnetic field with a peak field strength double that of the $(|\sigma_z|=1,|\ell|=0)$ case. Using these parameters, and assuming the laser is entirely absorbed, the peak ratio of the energy per unit length contained in the magnetic field to the laser energy per unit length is roughly $0.2\%$ for the CP $\ell=0$ and LP $|\ell|=1$ cases, slightly increasing for higher order $\ell$ modes.

The dependence of $\alpha_{abs}$ on $\ell$ is not currently understood for any absorption mechanism. At laser intensities of $I\lambda^2\geq10^{19}$ Wcm$^{-2} \mu$m$^2$ and pulse duration's greater than 100 fs, we can estimate the pump depletion length in underdense plasma ($n_e > 0.01 n_c$) as $L_{pd}=n_c c\tau/n_e$ where $\tau$ is the temporal FWHM \cite{Lu07,Decker96}. Rearranging, we obtain the absorption rate as, 
\begin{equation}\label{eq2.1}
    \alpha_{abs}=\frac{n_e}{2n_c c\tau}.    
\end{equation}
Given the absorption is independent of beam intensity and beam waist, we assume the model is approximately valid for low order OAM modes. If $\tau \gtrsim 100$ fs, and $n_e\gtrsim 0.01 n_c$, we can use Eq's. \ref{eq3} and \ref{eq2.1} to estimate the peak axial magnetic field strength of the $(|\sigma_z|=1,|\ell|=0)$ and $(\sigma_z=0, |\ell|=1)$ modes,
\begin{equation} \label{eq4}
    |B|_{max}\approx10 \frac{P[TW]\lambda^3[\mu m]}{w_0^4[\mu m]} kT.
\end{equation}

The fourth power dependence on the laser beam waist indicates a strong dependence on the beam f-number and relativistic self-focusing. Eq. \ref{eq4} should therefore be considered an under-estimate as the beam will self-focus beyond the diffraction limit giving rise to much stronger magnetic fields. Using Eq. \ref{eq4}, we can compare its prediction with the experimental results previously measured using CP Gaussian beams \cite{najmudin01}. In that experiment, the Vulcan laser ($40$ TW, $1$ ps, $\lambda=1.054\mu$m) was focussed into an underdense plasma where it was observed to self focus to a beam waist of $w_0\approx5 \mu$m. A peak magnetic field strength was observed of $\approx 700$ T, in good agreement with our Eq. \ref{eq4} estimate of $771$ T. The same experiment also measured an independence of the magnetic field strength on the plasma density, also in agreement with our model.  

Circularly polarizing large diameter, high power laser beams is not trivial requiring fragile and costly optics. However, with the new generation of PW class lasers coupled with off-axis spiral phase mirrors \cite{Longman:20}, axial magnetic fields on the order of 10's of kT may be feasible using LP OAM modes. Ultimately, the field strength will depend on the absorption rate of the laser which may diverge from our simple model when using ultra-high intensity beams $(I_0\gtrsim 10^{21}Wcm^{-2})$, high $|\ell|$ beams, and ultra-short pulse durations $(\tau<100fs)$. Additionally, realistic diffraction of OAM beams when generated by spiral phase optics is more complex leading to larger beam waists and lower intensities than one would expect using pure Gaussian beams \cite{Longman:17,Longman:20b}. While we do not include these modified beam waists and intensities here, it is simple to use the estimates from these previous works to estimate the reduction in the magnetic field strength. 

\begin{figure*} 
\includegraphics{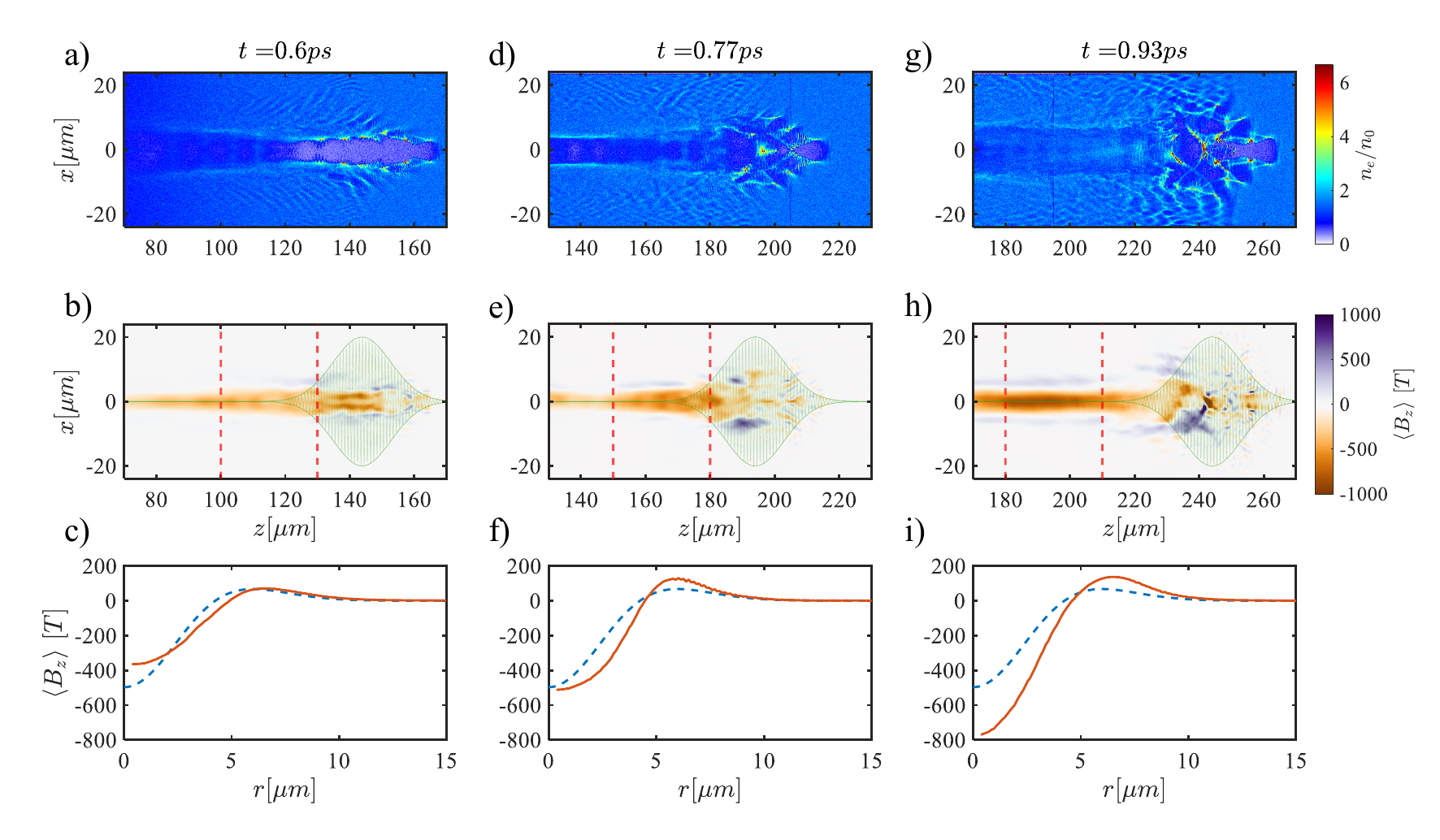}
\caption{\label{fig3} Simulation results for CP Gaussian driven magnetic fields at times (from left to right) $0.6$ ps, $0.77$ ps, and $0.93$ ps. Tiles a), d), and g): longitudinal electron density slice maps normalized to the initial electron density. Tiles b), e), and h): longitudinal axial magnetic field slice maps averaged over 33 fs, the laser pulse is overlayed in green and lineout regions are given by the red dashed lines. Tiles c), f), and i): Radial axial magnetic field lineouts averaged azimuthally, temporally, and longitudinally between the red dashed lines in the corresponding tile above, given by the red solid lines. The theoretical prediction of the model given by Eq. \ref{eq3} are shown by blue dashed lines. Laser-plasma parameters are given in the main text. 
}
\end{figure*}

\section{Simulation results}

To simulate the IFE, we use the 3D relativistic PIC code EPOCH \cite{Arber_2015}. We assume a fully ionized helium plasma with a super-Gaussian longitudinal shape to mimic the electron density from a gas jet given by, $n(z)=n_e \exp\{-[(z-350\mu m)/300\mu m]^{10}\}$ with initial electron and ion temperatures of 1 keV and 1 eV respectively. Using this profile, the plasma ramps from vacuum to $n_e$ over roughly 100 $\mu$m. The laser is polarized along the $\hat{x}$ axis, has a Gaussian temporal pulse shape, and is focused to the plane $z=135$ $\mu$m.

The laser pulse is tracked with a moving window 200 $\mu$m long and 48 $\mu$m $\times$ 48 $\mu$m in the transverse directions with grid cell sizes of 50 nm $\times$ 80 nm $\times$ 80 nm respectively. We use 4 particles per cell and have open boundaries throughout. After the magnetic field has been generated and before significant laser diffraction and filamentation occurs, the moving window stops at 1 ps and the evolution of the magnetic field is observed. Simulations were run for the $(\sigma_z=0,\ell=0)$, $(\sigma_z=1,\ell=0)$, $(\sigma_z=0,\ell=1)$ and $(\sigma_z=0,\ell=2)$ laser modes, with parameters $P=65$ TW, $\tau=100$ fs, $\lambda=1$ $\mu$m, $w_0=6$ $\mu$m, and $n_e=3\times 10^{19}$ cm$^{-3}$. A fifth simulation with $(\sigma_z=0,\ell=1)$, $P=50$ TW, $\tau=100$ fs, $\lambda=1$ $\mu$m, $w_0=7.2$ $\mu$m, and $n_e=3\times 10^{19}$ cm$^{-3}$ was also run to verify the decay model. 


\begin{figure*} 
\includegraphics{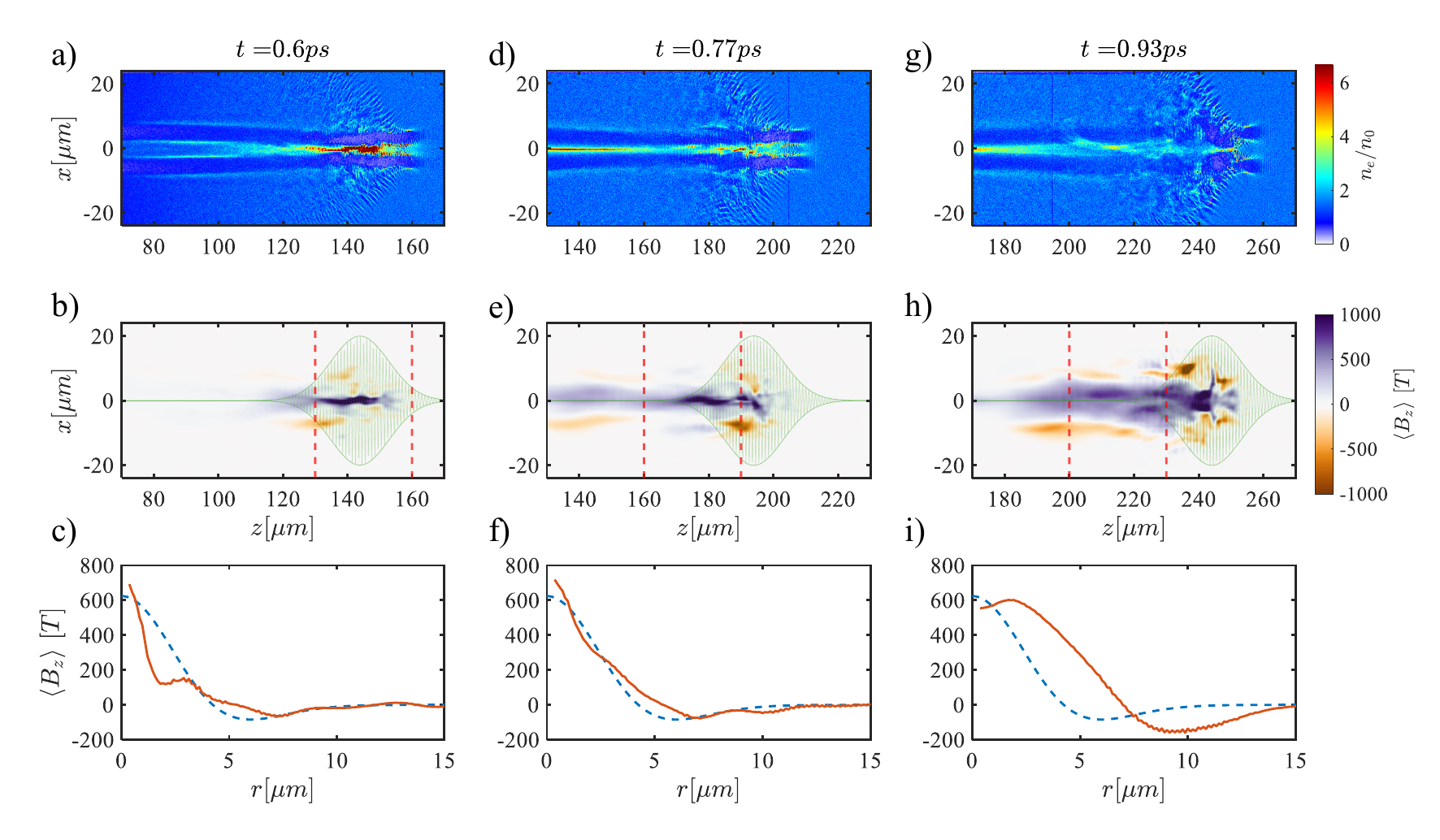}
\caption{\label{fig4} Simulation results for LP $\ell=1$ Laguerre Gaussian driven magnetic fields at times (from left to right) $0.6$ ps, $0.77$ ps, and $0.93$ ps. Tiles a), d), and g): longitudinal electron density slice maps normalized to the initial electron density. Tiles b), e), and h): longitudinal axial magnetic field slice maps averaged over 33 fs, the laser pulse is overlayed in green and lineout regions are given by the red dashed lines. Tiles c), f), and i): Radial axial magnetic field lineouts averaged azimuthally, temporally, and longitudinally between the red dashed lines in the corresponding tile above, given by the red solid lines. The theoretical prediction of the model given by Eq. \ref{eq3} are shown by blue dashed lines. Laser-plasma parameters are given in the main text.  
}
\end{figure*}

\begin{figure*} 
\includegraphics{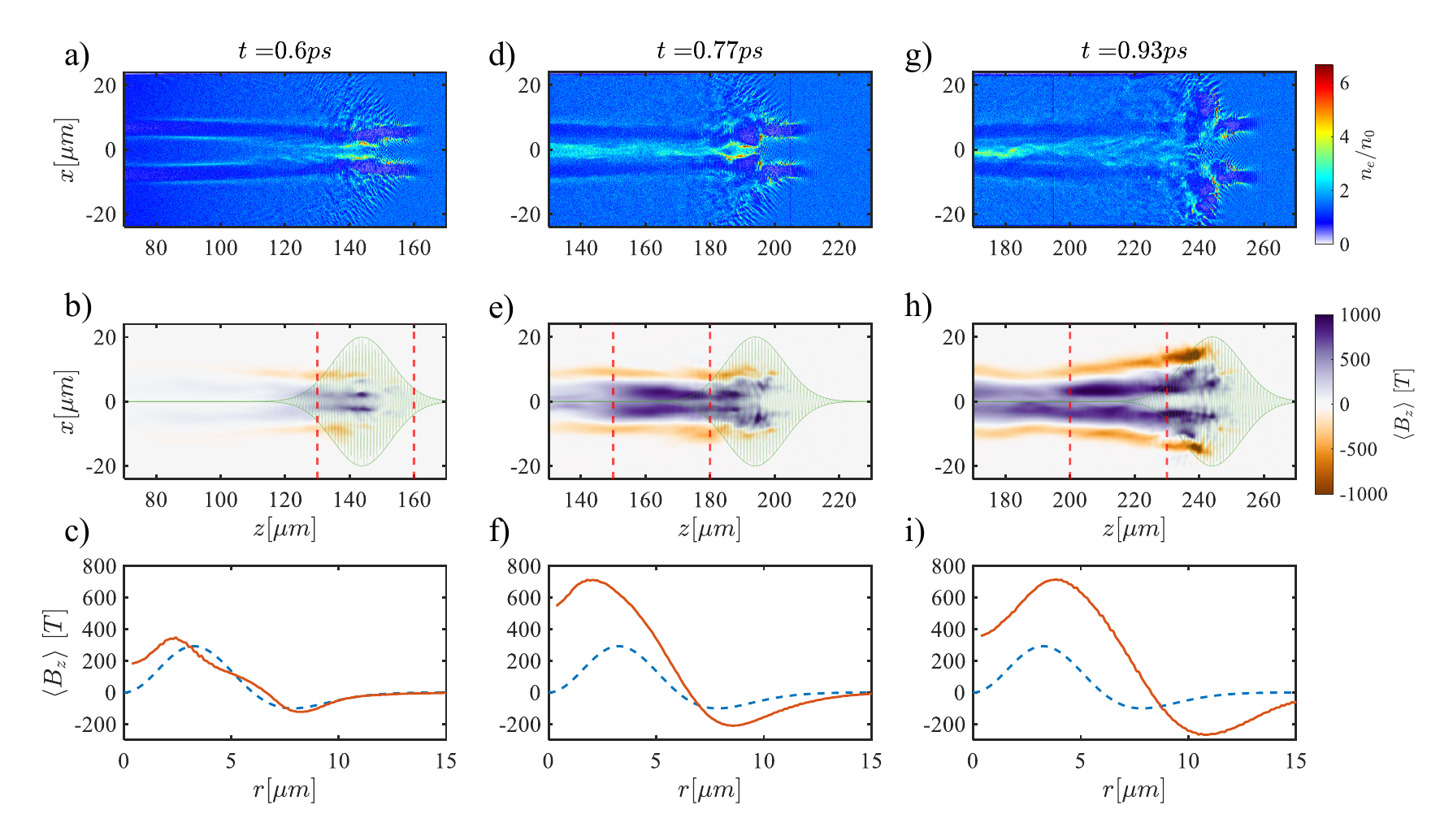}
\caption{\label{fig5} Simulation results for LP $\ell=2$ Laguerre Gaussian driven magnetic fields at times (from left to right) $0.6$ ps, $0.77$ ps, and $0.93$ ps. Tiles a), d), and g): longitudinal electron density slice maps normalized to the initial electron density. Tiles b), e), and h): longitudinal axial magnetic field slice maps averaged over 33 fs, the laser pulse is overlayed in green and lineout regions are given by the red dashed lines. Tiles c), f), and i): Radial axial magnetic field lineouts averaged azimuthally, temporally, and longitudinally between the red dashed lines in the corresponding tile above, given by the red solid lines. The theoretical prediction of the model given by Eq. \ref{eq3} are shown by blue dashed lines. Laser-plasma parameters are given in the main text.
}
\end{figure*}

\subsection{Circularly Polarized Gaussian Simulations}

A simulation was first run for the linearly polarized Gaussian beam to verify the null magnetic field result, and also to verify the absorption model. While we do not show the simulation results in this work the null field was verified, and the absorption rate of this beam was found to be 0.43 mm$^{-1}$ in excellent agreement with the predicted value of 0.435 mm$^{-1}$ from Eq. \ref{eq2.1}. A second identical simulation was run but with a circularly polarized Gaussian beam ($\sigma_z=1, \ell=0$) that measured absorption slightly enhanced to 0.48 mm$^{-1}$. 

Given the peak intensity of the CP Gaussian beam is $I_0=1.15\times10^{20}Wcm^{-2}$, the interaction with the plasma is strongly nonlinear with effects such as relativistic self-focussing, ponderomotive channeling, and filamentation strongly dominating the interaction. This can make it challenging to get clean magnetic field profiles out of the simulation, especially at later times when the boundary conditions start to have an effect on the simulation. Because of this, we examine the magnetic field at early times before these nonlinear effects have time to sufficiently grow. 

Fig. \ref{fig3} shows the axial magnetic field generation from the CP Gaussian mode at three early times in the simulation. Tiles a), d) and g) give the electron density normalized to the initial electron density at times $0.6$ ps, $0.77$ ps, and $0.93$ ps respectively. Below these tiles are transverse slices of the time-averaged axial magnetic fields (averaged over 33fs output dumps) relative to the laser pulse overlayed in green for each of the three times. The bottom three tiles give radial lineouts of the time averaged magnetic field that have been averaged azimuthally, temporally (over 33fs), and longitudinally between the red dashed lines in tiles b), e), and h). The simulation lineouts are given in red, whereas the theoretical prediction of Eq. \ref{eq3} is given with the blue dashed line. 

We find a remarkable overlap between the theory and the simulations at early times but later diverges from the theory as the laser begins to diffract, self-focus, and filament. In particular we see the peak magnetic field strength enhance beyond the vacuum theory at $t=0.93$ ps due to the self-focussing. We also note the persistence of the magnetic fields after the laser pulse has passed due to the residual magnetization of the plasma.

\subsection{Linearly Polarized OAM Simulations}

Next, simulations with LP OAM modes were run using both $\ell=1$ and $\ell=2$ modes, the results of which are given in Fig. \ref{fig4} and Fig. \ref{fig5} respectively with the same layout described for the CP Gaussian beams in Fig. \ref{fig3}. The absorption of the $\ell=1$ and $\ell=2$ modes were found to be 0.59 mm$^{-1}$ and 0.60 mm$^{-1}$ respectively. This increase in absorption may come from the larger cross section of the OAM modes, electron trapping within the donut mode leading to a more efficient laser-plasma energy transfer, or from a reduced ponderomotive channeling leading to a higher plasma density interacting and coupling with the laser directly.

Examining Fig. \ref{fig4} we see a good overlap of the simulated magnetic field and that predicted by the theory. Similar to the CP Gaussian simulations, we find the early times of the $\ell=1$ simulation agree well, but at later times as the laser starts to diffract, self-focus, and filament. In tile \ref{fig4}g) and h), we see the magnetic field expand radially outwards without compromising its peak strength. This surprising result could be related to the compressed electron density on axis giving rise to a higher azimuthal current, and a slow radial expansion outwards to fill the voids left by the ponderomotive force.  

For the $\ell=2$ simulation shown in Fig. \ref{fig5} we also see a good overlap between theory and simulation at early times, but at later times we find that the magnetic field amplifies to twice that of the predicted field, whilst still maintaining its shape. This is likely due to the more complex self-focussing of the LG modes as the donut mode both pinches into a tighter ring increasing intensity but maintaining the beam waist, and then collapsing to a smaller beam wasit at later times. During this self-focussing, additional angular momentum may be coupled to the plasma giving rise to the enhanced magnetic field. As our model does not include self-focussing, we do not expect to reproduce the results at later times. There is some infilling of the null within the magnetic field on axis as the plasma thermalizes, and then eventual radial expansion as the laser later diffracts. 

In the current configuration the plasma is radially isotropic and the axial magnetic field length is essentially limited to the Rayleigh length of the laser. For longer magnetic fields, one could use a preformed plasma channel to guide the laser and mitigate the effects of diffraction but is not explored in the current work. 

\begin{figure*} 
\includegraphics{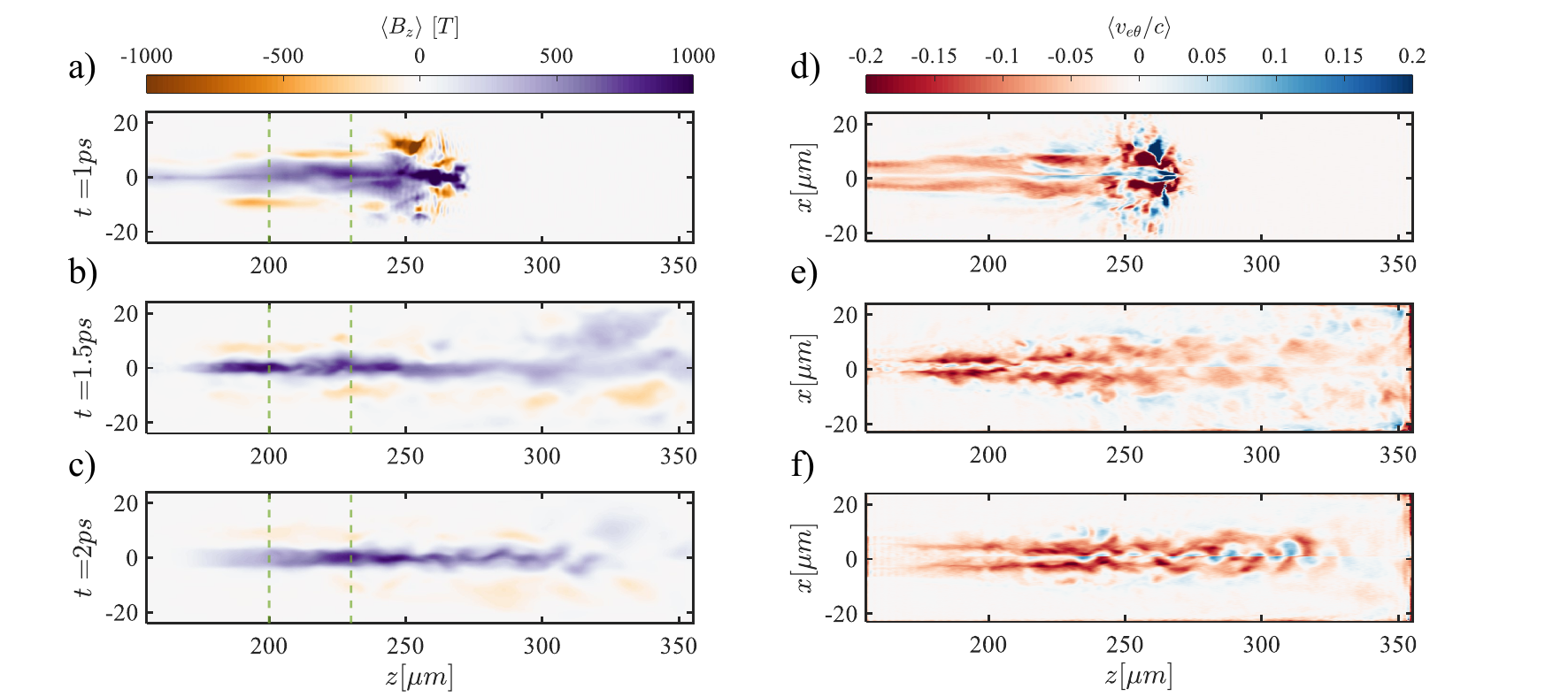}
\caption{\label{fig6} 2D longitudinal slices of the time averaged (averaged over 33fs output dumps) axial magnetic field and the corresponding time averaged azimuthal electron velocity for the $(\sigma_z=0, \ell=1)$ OAM mode for times $t=1,1.5,2$ ps. The green dashed lines indicate the regions used for sampling the field decay in Fig. \ref{fig7}.}
\end{figure*}

Fig. \ref{fig6}(a), (b) and (c) give the time averaged (averaged over 33fs output dumps) axial magnetic field profile at times 1 ps, 1.5 ps and 2 ps respectively for the $(\sigma_z=0, \ell=1)$ mode. We find the magnetic field extends the length of the simulation box (200 $\mu$m) approximately equal to the twice Rayleigh range ($z_0\approx 113$ $\mu$m).

At later times in Fig. \ref{fig6}(b) and (c), we find the $\ell=1$ mode to begin to pinch, kink, and twist into a 3D spring-like shape. Similar kinking is found for the $(|\sigma_z=1|,|\ell|=0)$ and $(|\sigma_z=0|,|\ell|=2)$ modes although not shown. Fig \ref{fig6} (d), (e), and (f) show transverse slices of the electron azimuthal velocity time-averaged over 33 fs. We see the azimuthal velocity map transition from a smooth cylindrical profile at 1 ps, to a more turbulent and kinked map at 2 ps. This could be related to the wobble instability found in theta-pinch configurations \cite{Freidberg78}. 



\section{Magnetic Field Decay}

The lifetimes of the magnetic fields driven by the IFE at relativistic intensities in collisionless plasmas are poorly understood. If we consider the plasma beta, $(\beta=2\mu_0 n_e k_BT/B^2)$ where $\mu_0$ is the permeability of free space, $k_B$ is Boltzmann's constant, and $T$ is the plasma temperature, we find that for magnetic field strengths of 1000 T, electrons with energies $T < 100$ keV are trapped in the magnetic field, while higher energy electrons are able to escape leading to the subsequent decay of the magnetic field. 

To model the decay, we consider the motion of the hot electrons in helical motion with both azimuthal and radial velocity components. From Fig. \ref{fig6} d) - f), we find the bulk of the hot electrons with OAM are born within a cylinder of radius approximately equal to the peak intensity radius of the OAM mode $r_{cyl}\approx w_0\sqrt{\ell/2}$. The higher order $|\ell|\geq2$ modes are more complex with coaxial fields and will not be considered in our model. 

The magnetic field as a function of time from a rotating cylinder of plasma can be derived from the Biot-Savart law \cite{Jackson:100964}. The current density can be given as $\bm{j}=q n_e(v_r \hat{r}+v_{\theta}\hat{\theta})$, the position vector as $\bm{r}(t)=(v_rt+r_0)\hat{r}+z\hat{k}$ where $r_0$ is the initial electron radius, and $\hat{r}$ is the rotating polar coordinate. Inserting these into the Biot-Savart equation we find, 
\begin{equation}\label{eq5}
    B(t)=\frac{\mu_0 e n_e}{2}\int_0^{c\tau} dz\int_0^{r_{cyl}}\frac{v_\theta(v_r t+r_0)r_0dr_0}{\left[(v_rt+r_0)^2+z^2\right]^{3/2}}.
\end{equation}
We note that the axial length of the plasma for integration is restricted to approximately one pulse length $c\tau$.
As the plasma is collisionless, we can assume the angular momentum acquired by the electrons is in steady state, that is $L_z=m_e v_\theta r_0$ is constant. Solving Eq. \ref{eq5} is straightforward if we perform the radial integral first yielding the result, 
\begin{equation}\label{eq9}
    B(t)=B_0\left[\sinh^{-1}\left(\frac{c\tau}{v_rt}\right)-\sinh^{-1}\left(\frac{c\tau}{v_rt+r_{cyl}}\right)\right].
\end{equation}
Here, $B_0$ is the peak axial magnetic field at genesis.

To estimate the radial velocity of the electrons we can use the gradient of the plasma pressure \cite{haines01}. To first order, the pressure can be attributed solely to the thermal pressure given by,
\begin{equation}\label{eq10}
    \frac{n_e m_i}{Z}\frac{dv_r}{dt}=-\frac{\partial}{\partial r}\left(\frac{2}{3}\alpha_{abs}\int I(t)dt\right).
\end{equation}
Here, $m_i$ and $Z$ are the ion mass and ionization state respectively. By using the ion mass, we assume the plasma to be quasineutral which was previously shown to be approximately valid on these timescales and laser intensities \cite{haines01}. Integrating the laser temporal function is trivial as shown earlier in the paper leaving a second order ODE,
\begin{equation}\label{eq11}
    \frac{d^2r}{dt^2}+\frac{Z\alpha_{abs}I_0\tau}{2n_em_i}\frac{\partial}{\partial r}|\psi_{\ell}|^2=0.
\end{equation}
There are no known analytic solutions to Eq. \ref{eq11} and we opt instead to use a parabolic approximation of $\psi_\ell$ given by, 
\begin{equation}\label{eq12}
    |\psi_\ell|^2\approx A-\left(\frac{r}{w_0}-\sqrt{\frac{\ell}{2}}\right)^2,
\end{equation}
where $A$ is an arbitrary amplitude scaling variable that is lost when we use the $\partial/\partial r$ operator.

The soltuion of the ODE is elementary using the parabolic approximation. Averaging the radial velocity over all positions within a beam waist, we find
\begin{equation}\label{eq13}
    \langle v_r\rangle =\sqrt{\frac{Z\alpha_{abs}I_0\tau}{4\exp(2) n_em_i}}.
\end{equation}
Substituting our previous simulation values, we find an average radial velocity of $\langle v_r\rangle\approx 0.01c$. Given this, we can use the approximation that $(c\tau)^2\gg(v_r t)^2$ given the timescale of the magnetic field decay is observed to be on the order of a picosecond, allowing us to simplify Eq. \ref{eq9} to
\begin{equation}\label{eq6}
    \frac{B(t-t_0)}{B_0}\approx \ln\left(1+\frac{1}{t/t_D + 0.582}\right).
\end{equation}
Here $t_D=w_0/(v_{er}\sqrt{2})$ is the decay parameter for the $|\ell|=1$ mode, and $t_0$ is the genesis time. The factor of 0.582 is set such that $B(0)=B_0$. Combining the results of Eq. \ref{eq13} and Eq. \ref{eq6} allows us to derive the scaling law for the time decay parameter of the $|\ell|=1$ driven magnetic fields,
\begin{equation}\label{eq15}
    t_D[ps]\approx\frac{1}{8}\frac{w_0^2[\mu m]}{\lambda [\mu m]}\sqrt{\frac{A}{Z}}\frac{1}{\sqrt{P[TW]}},
\end{equation}
where $A$ is the atomic mass number of the plasma. While this model is not valid for $|\ell|\geq2$, it could be suitable for also modelling the CP $\ell=0$ mode given its similarities to the $|\ell|=1$ magnetic field profile. 

\begin{figure} 
\includegraphics{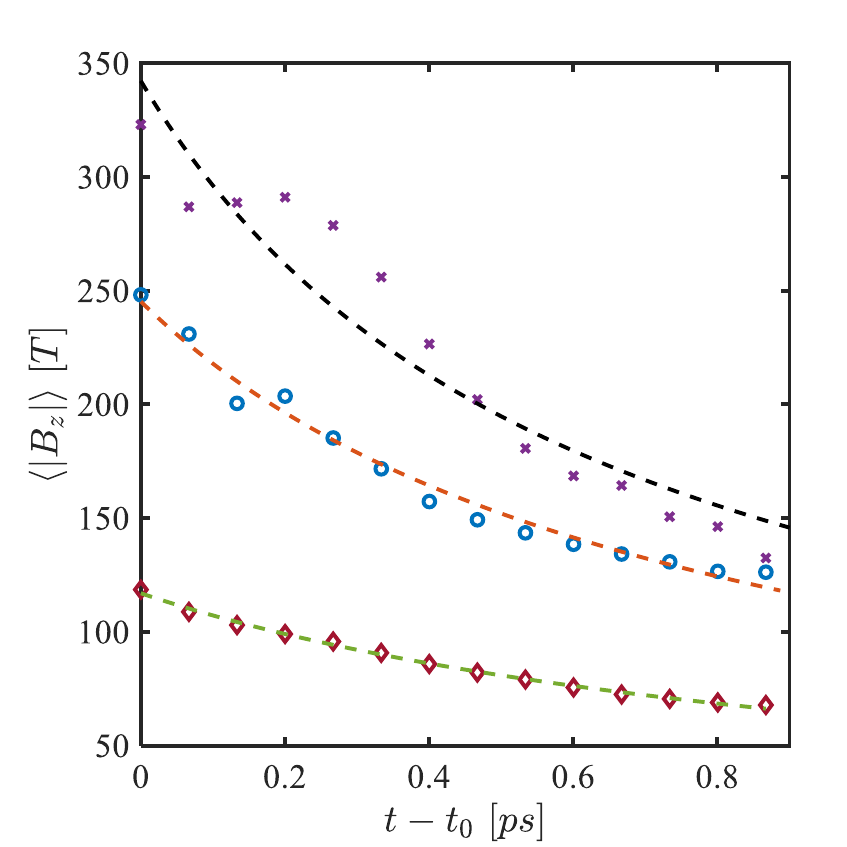}
\caption{\label{fig7} Average magnetic field within a cylinder of radius $r_{cyl}=w_0/ \sqrt{2}$ and length 30 $\mu$m from $z-200-230\mu$m as a function of time. The LP $65$ TW, $\ell=1,w_0=6$ $\mu$m mode is given by the blue circles, and fitted with the red dashed line with decay parameter $t_D=0.86$ ps. The LP $50$ TW, $\ell=1,w_0=7.2$ $\mu$m mode is given by the red diamonds, and fitted with the green dashed line with decay parameter $t_D=1.18$ ps. The $65$ TW, $\sigma_z=1,\ell=0,w_0=6$ $\mu$m mode is given by the purple crosses, and fitted with the black dashed line with decay parameter $t_D=0.69$ ps.}
\end{figure}

Using our simulation values, we estimate the decay parameter as $t_D=0.8$ ps. Determining the decay time numerically is challenging due to the magnetic field instabilities, as well as laser self-focussing causing the magnetic field to be dynamic on axis. We therefore radially, azimuthally, and longitudinally average the magnetic field through a cylinder of radius $r_{cyl}=w_0/\sqrt{2}$ and length $c\tau$. The longitudinal region is indicated by the dashed green bars from $z=200-230$ $\mu$m as shown in Fig. \ref{fig6}. The average value of the magnetic field in this region is plotted as a function of time in Fig. \ref{fig7} for three simulations. The simulation for the $65$ TW, $(\sigma_z=0, \ell=1),w_0=6$ $\mu$m mode is shown with blue circles where as the $50$ TW, $(\sigma_z=0, \ell=1)$ and $w_0=7.2$ $\mu$m simulation is shown with red diamonds. The purple crosses represent the data from the initial CP Gaussian $(\sigma_z=1, \ell=0)$ simulation with $P=65$ TW, $\ell=0,w_0=6$ $\mu$m.  

Using a nonlinear least squares fit of Eq. \ref{eq6} to the data, we find a numerical decay parameter of $0.86$ ps for the $\sigma_z=0, \ell=1,w_0=6$ $\mu$m mode, shown by the red dashed line in Fig. \ref{fig7}. This is in very good agreement with the analytic estimate of $0.8$ ps. Considering the $\sigma_z=0,\ell=1,w_0=7.2$ $\mu$m case with a laser power of $50$ TW, we predict a decay parameter of $\tau=1.3$ ps and numerically find a value of $1.18$ ps shown by the green dashed line in Fig. \ref{fig7}. 

Finally for the $(\sigma_z=1, \ell=0)$ case we find a predicted decay also of $0.8$ ps, and numerically find a decay time of 0.69 ps as shown by the black dashed line. This reduced decay time could be due to the stronger self-focussing of the CP Gaussian mode as it has a higher peak intensity than the LG modes. The increased self-focussing results in a smaller beam waist, and hence a shorter decay time.

Given the dependence of the decay time on the laser beam waist, one could control the magnetic field lifetime by adjusting the f-number to higher values at a cost of magnetic field strength. An optimal f-number is therefore needed to balance the magnetic field strength, axial length, and decay time to suit a given application of interest. Alternatively one could adjust the decay time of the magnetic field by changing the plasma to a heavier ion species. 

\section{Summary}
In summary, we have successfully demonstrated and characterized the IFE with CP Gaussian and linearly polarized OAM modes over large spatial and temporal scales. We have derived a description of OAM driven magnetic fields in underdense plasma and introduced a model for the subsequent decay of the magnetic field. Magnetic fields with strengths up to 1 kT extending upto $200$ $\mu$m and persisting for several picoseconds have been demonstrated. Simulations indicate an increased absorption rate when using increased OAM mode numbers. The decay time, axial length, and magnetic field strength are all functions of the laser beam waist and can be optimized to suit an experiment as such. Plasma channel guiding of the laser pulse could extend the magnetic field over several Rayleigh lengths and would instead be limited to the pump depletion length. 

With the recent demonstration of high intensity OAM modes in high-power laser facilities with parameters similar to those assumed in this work, experimental verification of the magnetic fields could now be feasible \cite{Longman:20}. Generation and control of multi-kT magnetic fields will open up new opportunities in a number of areas of high energy density physics and laboratory astrophysics including particle acceleration schemes, magnetic re-connection, fast ignition and fundamental physics.

\begin{acknowledgments}
This work was performed under the auspices of the U.S. Department of Energy by Lawrence Livermore National Laboratory under Contract DE-AC52-07NA27344, supported by the Natural Sciences and Engineering Research Council of Canada (RGPIN-2019-05013), and support provided by WestGrid and Compute Canada. This work was in part funded by the UK EPSRC grants EP/G054950/1, EP/G056803/1, EP/G055165/1 and EP/ M022463/1. The authors would like to thank and acknowledge J. Myatt, J. Ludwig, and P. Michel for helpful discussions.
\end{acknowledgments}


\providecommand{\noopsort}[1]{}\providecommand{\singleletter}[1]{#1}%

\end{document}